\documentclass[aps,prl,twocolumn,superscriptaddress,showkeys,citeautoscript]{revtex4-1}

\usepackage[colorlinks=true,bookmarks=false,citecolor=blue,linkcolor=red,hyperfootnotes=true,urlcolor=blue]{hyperref}

\usepackage{physics}
\usepackage{graphicx}
\usepackage{dcolumn}
\usepackage{bm}
\usepackage[dvipsnames]{xcolor}
\usepackage{wasysym}
\usepackage{lipsum}
\usepackage{enumitem}
\usepackage{booktabs} 
\usepackage{gensymb}
\usepackage{xcolor}
\usepackage{amssymb}
\usepackage{booktabs}
\usepackage{array}
\newcolumntype{L}{>{$}l<{$}}
\newcolumntype{C}{>{$}c<{$}}
\newcolumntype{R}{>{$}r<{$}}

\usepackage{mathtools}

\usepackage{blindtext}

\makeatletter
\def\p@subsection{}
\makeatother

\newcommand{\code}[1]{\texttt{#1}}
\newcommand{\ii}{\code{i}}

\begin{document}

\title{Bond-Peierls polaron: Moderate mass enhancement and current-carrying ground state}

\author{Matthew R. Carbone} \email{mcarbone@bnl.gov}
\affiliation{Computational Science Initiative, Brookhaven National Laboratory, Upton, New York 11973}
\affiliation{Department of Chemistry, Columbia University, New York,
New York 10027}

\author{Andrew J. Millis} \email{ajm2021@columbia.edu}
\affiliation{Department of Physics, Columbia University, New York,
New York 10027} 
\affiliation{Center for Computational Quantum Physics, Flatiron Institute, 162 5$^{th}$ Avenue, New York, NY 10010}

\author{David R. Reichman} \email{drr2103@columbia.edu}
\affiliation{Department of Chemistry, Columbia University, New York,
New York 10027}

\author{John Sous} \email{js5530@columbia.edu}
\affiliation{Department of Physics, Columbia University, New York,
New York 10027}

\date{\today}

\begin{abstract}
We study polarons in the one-dimensional Bond-Peierls electron-phonon model, in which phonons on bonds of a lattice modulate the hopping of electrons between lattice sites, and contrast the results to those known for the Breathing-Mode Peierls problem.  By inspecting the atomic limit, we show that polaronic dressing and mass enhancement of Bond-Peierls polarons depend on the momentum dependence of the phonons. For dispersionless phonons, Bond-Peierls polarons are perfectly localized in the atomic limit, unlike their Breathing-mode counterparts, because  the carrier creates  a string of phonon excitations that can only be annihilated via processes that retrace the carrier to its original site. However, inclusion of phonon dispersion leads to a non-divergent polaron mass even in the atomic limit and depending on the form of the phonon dispersion may lead to a transition to a non-zero-momentum ground state akin to that found in the Breathing-mode Peierls model.
\end{abstract}


\maketitle

\emph{Introduction}.--- It has been appreciated for well over fifty years that the lattice distortions of a solid can strongly modulate electronic dynamics, inducing fluctuations in both the electron's potential and kinetic energies. Much of our generic understanding of this phenomenon has relied on simple models, such as the Holstein~\cite{holstein1959studiesI,holstein1959studiesII} and Fr\"ohlich~\cite{frohlich1950xx,frohlich1954electrons} models, that focus on the changes to the electronic potential energy and neglect the lattice-induced modulation of hopping integrals. These models, which we refer to generically as Holstein models, provide what are now standard ``textbook" results for the behavior of polarons~\cite{alexandrov1994bipolarons}. In recent years, investigations of models where phonons modulate the electron's hopping amplitude (kinetic energy) \cite{BarisicPeierls1,BarisicPeierls2,BarisicPeierls3,su1979solitons}, which we refer to generically as Peierls models, have revealed much richer physics including a polaronic transition  as a function of the coupling at which the ground state momentum becomes non-zero and a  dependence of the mass on coupling strength that changes from a regime of mass enhancement to one of mass reduction after the transition where the mass diverges \cite{marchand2010sharp}. This example provides direct evidence of a sharp change in polaronic behavior driven by the electron-phonon coupling, a phenomenon long debated in the literature~\cite{feynman1955slow}. 

The different behavior of polarons in Holstein versus Peierls models  presents fundamental questions.  In quantum materials polarons and bipolarons may contribute to the stabilization of emergent collective phases such as superconductivity. In traditional models such as the Holstein model, bipolarons become very heavy since phonons mediate a local effective attraction between already heavy polarons~\cite{bonvca1999holstein,BoncaHBipolaron}. In contrast, in models in which a breathing lattice distortion modulates the electron's hopping, phonons induce an effective pair-hopping interaction that binds polarons without sacrificing their kinetic energy, forming strongly bound yet light-mass bipolarons. Such bipolarons can be several orders of magnitude lighter than their counterparts in the Holstein model, even in regimes of similar electron, phonon and interaction energy scales~\cite{SousBipolaron}. Light bipolarons open the door to a new pathway to phonon-mediated high-temperature superconductivity~\cite{CRFBipolaron,SousBipolaron}. This calls for deeper investigation of general and model-specific behavior of polaronic mass enhancement in systems in which the lattice couples to the electron's kinetic energy. 

Two generic classes of Peierls models exist: the more widely studied ``Breathing-Mode'' Peierls (BMP) model in which the hopping across a bond (link)  is modulated by the difference in phonon amplitudes at opposite ends of the bond, $\hat{X}_{i} - \hat{X}_{j}$, and the Bond-Peierls (BP) model in which phonons that reside on the bonds, $\hat{X}_{i,j}$ ($i, j$ labels the bond between sites $i$ and $j$), modulate the electron hopping~\cite{kagan1976role}.  In this work, we focus on the less studied BP model, contrasting our findings with known results \cite{marchand2010sharp} (which we reproduce) for the BMP model.

Using the recently introduced numerically exact Generalized Green's Function Cluster Expansion (GGCE) approach~\cite{CarboneGGCE}, we compute exact one-particle Green's functions of polarons in the Bond-Peierls model at $T=0$. Our main observations and conclusions are as follows:
\begin{enumerate}[leftmargin=*]
    \item  For dispersionless phonons the effective mass of the Bond-Peierls polaron increases monotonically with the coupling strength. This behavior differs dramatically from that of Breathing-Mode Peierls polarons, which exhibit a transition as a function of the coupling strength at a critical coupling $\lambda_\mathrm{c}$  from a weak-coupling regime where the ground state has zero momentum to a strong-coupling regime at which the ground state has a non-zero momentum; the transition also separates a regime of mass enhancement as the coupling is increased from a regime of mass reduction.
    \item The polaronic dressing in the Bond-Peierls model depends strongly on the phonon dispersion. For strictly dispersionless phonons, there is perfect localization of the polaron in the atomic limit. Viewed complementarily, this corresponds to a  geometric process (absent in the Breathing-Mode Peierls and Holstein models) in which the carrier creates phonon configurations in such a way that it must retrace its path back to the original site in order to fully annihilate the phonons.
    \item Inclusion of phonon dispersion in the Bond-Peierls model restores a non-vanishing carrier velocity even in the $t=0$ limit and allows for the possibility of a transition to a state with non-zero ground state momentum. 
\end{enumerate}
We argue that these results do not depend strongly on spatial dimensionality. Thus, we expect that strong-coupling properties of polarons and bipolarons of Peierls models found in one dimension will qualitatively carry over to higher dimensions.

 \begin{figure}
    \includegraphics[width=\columnwidth]{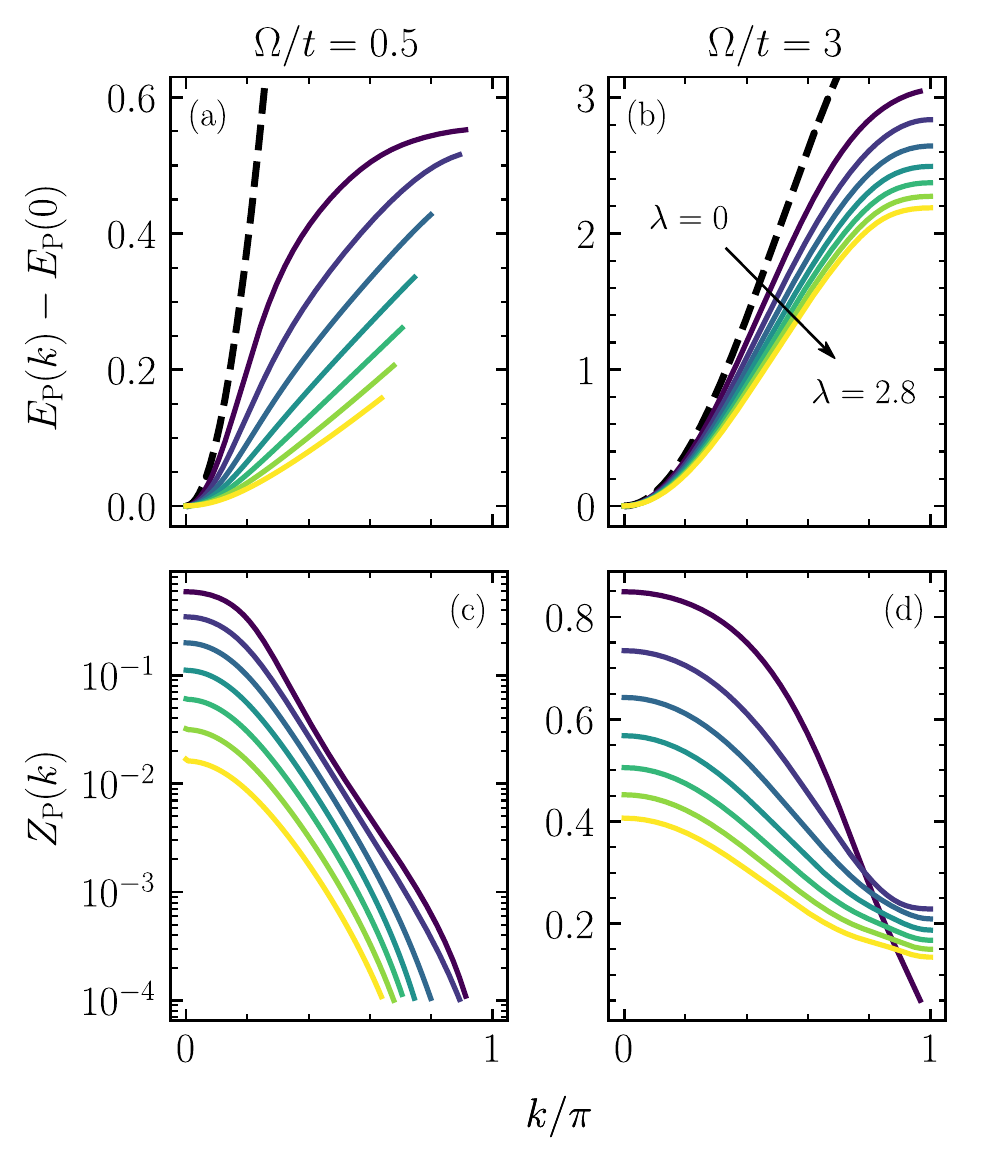}
    \caption{Polaron dispersion $E_\mathrm{P}(k) - E_{\rm P}(0)$ (a), (b) and quasiparticle weight $Z_{\rm P} (k)$ (c), (d) in the Bond-Peierls model with dispersionless phonons as a function of $\lambda \in \{0.4, 0.8, 1.2, 1.6, 2, 2.4, 2.8\}$ (from top to bottom) for two different adiabaticity parameters $\gamma = 1/16$ and $3/4$. The free-particle ($\lambda=0$) dispersion is shown as a dashed black line. The non-monotonic evolution of $Z_\mathrm{p}(k)$ with $\lambda$ and $k$  in (d) can be understood in terms of an avoided crossing of the polaron band with a one-phonon state at energy $\Omega$ relative to the  band minimum ($k=0$), which occurs for small $\lambda$. Here, what is plotted is $Z$ for the lowest state at a given $k$. For small $\lambda$ and at $k$ beyond the crossing point, this branch is almost entirely of phonon character.  As $\lambda$ increases the avoided crossing wavevector moves to the edge of the zone, and for $\lambda \gtrsim 0.5$ this physics becomes irrelevant.}
    \label{fig:fig1}
\end{figure}

\emph{Model and Methods}.---  We consider the Bond-Peierls model in one dimension (1D), defined as 
\begin{multline} \label{eq: model}
    \hat{{\mathcal H}} = -t \sum_{i} (\hat{c}_i^\dagger \hat{c}_{i+1} + {\rm h.c.}) +  \frac{1}{N}\sum_{ij} \Omega_{i-j}\hat{b}_{i+1/2}^\dagger \hat{b}_{j+1/2} \\
    + \alpha\sum_{i} (\hat{c}_i^\dagger \hat{c}_{i+1} + {\rm h.c.})  (\hat{b}_{i+1/2}^\dagger + \hat{b}_{i+1/2}).
\end{multline}
 Here, $t$ is the hopping amplitude for a carrier (spin is irrelevant here as we consider only one particle) that lives on the sites $i$ of the lattice (with lattice constant $a=1$) and is described by fermion creation (annihilation) operators $\hat{c}^\dagger$ ($\hat{c}$), $\Omega_{i-j}$ ($\hbar = 1$)  gives the dispersion of optical phonons modeled as Einstein oscillators located on the bonds of the lattice ($i+1/2$) and described by boson operators $\hat{b}^\dagger$ and $\hat{b}$, and $\alpha$ represents the electron-phonon coupling constant into which we have absorbed some constant parameters (e.g. the oscillator mass, etc.). To characterize the effects of the interaction, we define two dimensionless parameters: $\lambda = \alpha^2/\Omega t$, the dimensionless coupling, characterizes the strength of the interaction ($\lambda$ is the ratio of the ground state (GS) energy in the atomic limit to that in the free electron limit) and $\gamma = \Omega / {\mathcal W}$ (where ${\mathcal W}$ is the free electron bandwidth, ${\mathcal W}=4t$ in 1D), the adiabaticity parameter, specifies the degree of competition between distinct energy scales.

We use the GGCE approach~\cite{CarboneGGCE} to compute the carrier Green's function, $G(k, \omega) = \mel{0}{\hat{c}_k \hat{G}(\omega) \hat{c}_k^\dagger}{0}$, where $\hat{G}(\omega) = \left[\omega - H + \ii \eta\right]^{-1}$ is the propagator, with artificial broadening $\eta.$ The GGCE is a systematic, numerically exact expansion of the equation of motion of the propagator in the spatial extent of phonon clusters of size $M$ sites, and total phonon number $N$. A value of $M_{\rm max} = 4$ and $N_{\rm max}= 14$ allows us to converge all ground state results presented here to the exact limit.

\begin{figure}
    \includegraphics[width=\columnwidth]{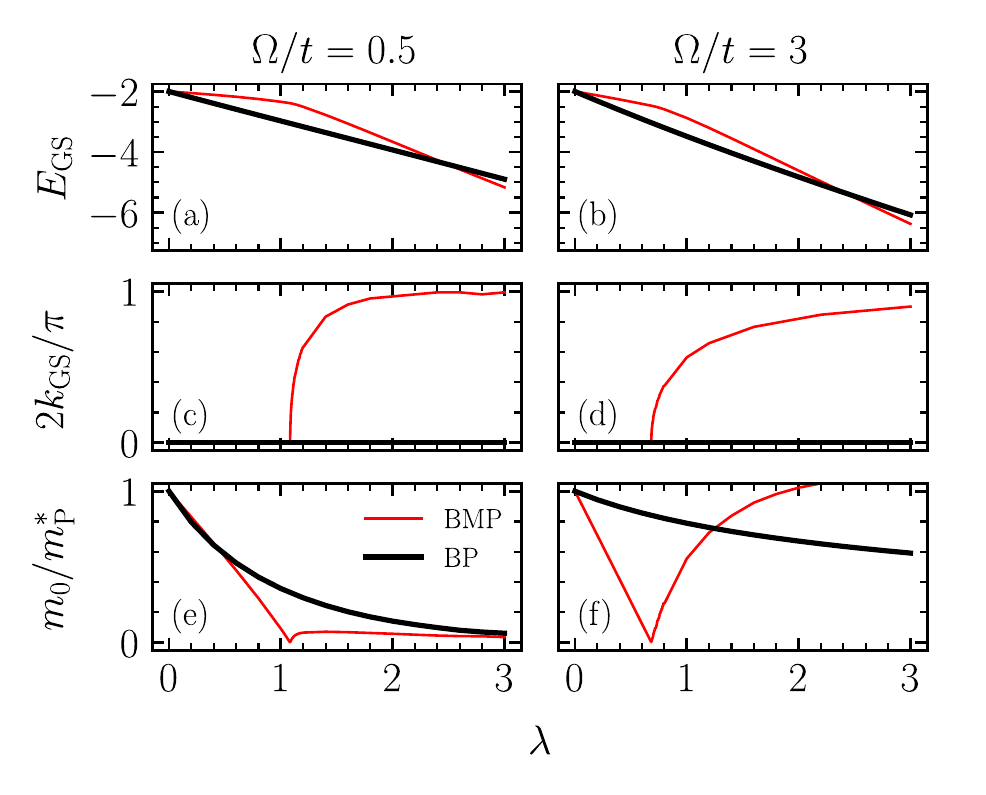}
    \caption{Ground state (GS) properties and polaron effective mass $m_\mathrm{P}^*$ in the Bond-Peierls (BP) model with dispersionless phonons versus that in the Breathing-Mode Peierls (BMP) model  as a function of $\lambda$  for $\gamma = 1/16$ and $3/4$. $m_0 = 1/2t$ is the free electron mass. $\lambda = \alpha^2/\Omega t$ for the Bond-Peierls model, and $\lambda = 2\alpha^2/\Omega t $ for the Breathing-Mode Peierls model.}
    \label{fig:fig2}
\end{figure}

\begin{figure*}
    \includegraphics[width=\textwidth]{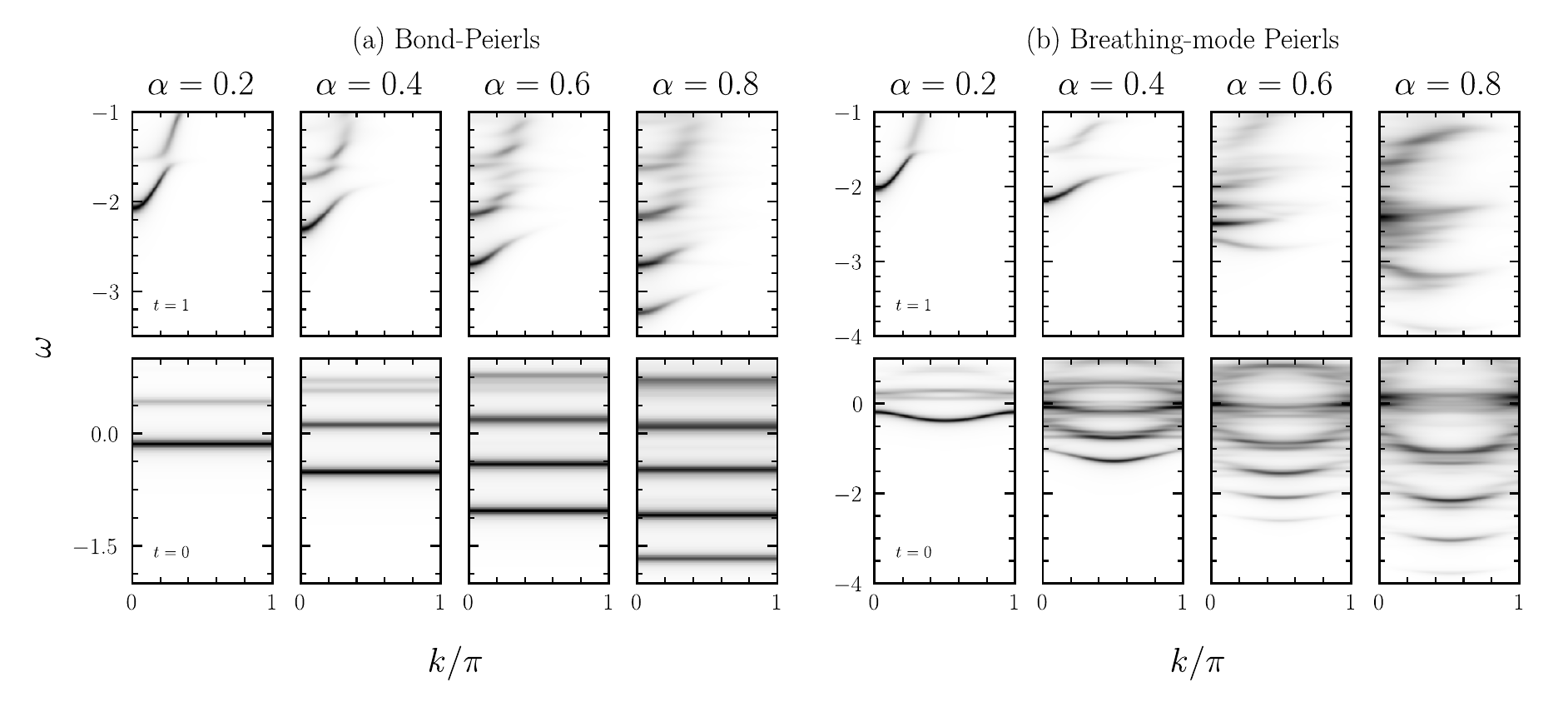}
    \caption{Spectral functions for $\Omega = 0.5$ and broadening factor $\eta = 0.04$
    of the (a) Bond-  and (b) Breathing-Mode Peierls models with dispersionless phonons  at $t=1$ (top row) and $t=0$ (bottom row) for a variety of different coupling strengths $\alpha.$ 
    }
    \label{fig:fig3}
\end{figure*}

\emph{Results}.--- A polaron forms when a discrete peak in the spectral function appears below the free electron energy, $-2t$ (in 1D).  We find that a polaron forms for any $\lambda>0$, see Figs.~\ref{fig:fig1} and~\ref{fig:fig2}. We track in Fig.~\ref{fig:fig1}(a) and (b) the polaron dispersion, $E_{\rm P}(k)$, as a function of $\lambda$ for two adiabaticity parameters $\gamma = 1/8$, $3/4$, interpolating between the deep and intermediate adiabatic regimes. For very small couplings, the polaron band exhibits continuity from the free electron band and a cosine-like form. As the coupling strength increases the bandwidth drops, driven by shifts at higher $k$ to lower energies, and a flattening of the band minimum near $k=0$. This behavior, reminiscent of typical polaronic dressing, grows monotonically with larger coupling strength and in stronger adiabaticity regimes. This picture is corroborated in Fig.~\ref{fig:fig1} (c) and (d) showing an increasingly small quasiparticle weight, $Z_{\rm P} (k) = \abs{\bra{{\rm GS}_{\rm P}} {c_{k}^\dagger \ket{0}}}^2$ ($\ket{{\rm GS}_{\rm P}}$ is the GS polaron wavefunction, and $\ket{0}$ is the zero-carrier, zero-phonon vacuum state), with increasing couplings and for large $k$. The behavior at large $k$ results from the proximity of the polaron at this $k$ to the polaron $+$ one-phonon continuum, $E_{\rm GS} + \Omega$, also seen in other models~\cite{bonvca1999holstein,marchand2010sharp}.

Figure~\ref{fig:fig2} contrasts the evolution of polaron properties with $\lambda$ in the Bond-Peierls model with its Breathing-Mode Peierls counterpart in which the electron-phonon coupling
is replaced by
\begin{equation}
    \hat{V}^{\rm BMP}_{\rm e-ph} = \alpha \sum_{i} (\hat{c}_i^\dagger \hat{c}_{i+1} + {\rm h.c.})  (\hat{b}_{i}^\dagger + \hat{b}_{i} - \hat{b}_{i+1}^\dagger - \hat{b}_{i+1})
\end{equation}
and phonons on the sites instead of the bonds. The Breathing-Mode Peierls polaron exhibits a sharp transition as a function of $\lambda$ from one with a single minimum at $k=0$ for $\lambda < \lambda_\mathrm{c}$ to one with two degenerate minima at  $\pm k \neq 0 $ at $\lambda>\lambda_\mathrm{c}$~\cite{marchand2010sharp}.  This transition does not occur in the Bond-Peierls model with dispersionless phonons. The sharp transition of the Breathing-Mode Peierls polarons manifests as a sharp divergence in $m_\mathrm{P}^* = \left[\partial^2 E_{\rm P}(k) / \partial k^2 |_{k=k_\mathrm{GS}} \right]^{-1}$ at $\lambda_\mathrm{c}$ before which the mass grows with $\lambda$ and after which the mass decreases with $\lambda$.  In contrast, we see no evidence of a transition or the associated  divergence in $m_\mathrm{P}^* (\lambda)$ of the Bond-Peierls polarons with dispersionless phonons.  Instead, we find the typical behavior in which the polaron mass increases  with increasing $\lambda$, however, with a dependence clearly weaker than exponential, unlike in the Holstein model.  This translates into relatively lighter masses relative to those in the Holstein model even for very large (and unrealistic) couplings such as $\lambda = 3$ in the adiabatic limit $\Omega = 0.5t$, where the Bond-Peierls polaron mass $\sim 10 m_0$ remains several orders of magnitude lighter than the Holstein polaron mass $\sim 6 \times 10^4 m_0$~\cite{goodvin2006green}.

The mechanism behind polaron formation and mass enhancement in the Bond-Peierls and Breathing-Mode Peierls models becomes transparent in the atomic limit $t=0.$ Up to a constant shift in energy, the Hamiltonian of a generic linearly coupled electron-phonon system in the atomic limit reads
\begin{equation}\label{Eq:atomiclimit}
\hat{\mathcal{H}} = \frac{1}{2} \sum_i \hat{P}_i^2 + \frac{1}{2} \sum_{i,j} \hat{X}_i \mathcal{D}_{i,j} \hat{X}_j + \sum_i \hat{X}_i \hat{O}^{\rm e}_i.
\end{equation}
Here, we have recast the Hamiltonian in the basis of first-quantized oscillator operators (oscillator mass $M=1$), $\hat{P}_i = \ii\sqrt{\frac{\Omega}{2}}(\hat{b}_i^\dagger - \hat{b}_i)$ and $\hat{X}_i = \sqrt{\frac{1}{2\Omega}} (\hat{b}_i^\dagger + \hat{b}_i)$, and $\mathcal{D}_{i,j} = \mathcal{D}_0 \delta_{i,j} + \mathcal{D}_1 ( \delta_{j,i+1} + \delta_{j,i-1})$ is the Hessian matrix with $\mathcal{D}_0 = \Omega^2$, a local harmonic term and $\mathcal{D}_{i\neq j}$ provides phonon dispersion, and $\hat{O}^{\rm e}_i$ is an electronic operator that couples linearly to the oscillator, with strength $\tilde{\alpha} = \alpha \sqrt{2\Omega}$. In any Hamiltonian of the form of Eq.~\ref{Eq:atomiclimit} the electron and phonon coordinates can be decoupled via a transformation in which  the oscillator coordinate is displaced $\hat{X}_i \rightarrow \hat{\tilde{X}}_i =\hat{X}_i + \sum_j\mathcal{D}^{-1}_{i,j} \hat{O}^{\rm e}_j$ to yield
\begin{equation}\label{Eq:Diagatomiclimit}
\hat{\mathcal{H}} = \frac{1}{2} \sum_i \hat{P}_i^2 + \frac{1}{2} \sum_{i,j}   \hat{\tilde{X}}_i \mathcal{D}_{i,j} \hat{\tilde{X}}_j - \frac{1}{2} \sum_i  \hat{O}^{\rm e}_i \mathcal{D}^{-1}_{i,j} \hat{O}^{\rm e}_j.
\end{equation}
 In the Bond-Peierls model $\hat{O}^{\rm e}_i = \tilde{\alpha} \hat{O}^{\rm BP}_i =\tilde{\alpha} (\hat{c}_{i+1/2}^\dagger \hat{c}_{i-1/2} + \hat{c}_{i-1/2}^\dagger \hat{c}_{i+1/2})$ and a simple calculation shows that in the absence of phonon dispersion ($D_{i,j}=D_0\delta_{ij}$),  the electronic ground state is strictly localized in the atomic limit.  Dispersion arises from $\mathcal{D}^{-1}_{i,i\pm 1}$, which gives rise to a second-neighbor hopping of the carrier, and depending on the sign of $\mathcal{D}^{-1}_{i,i\pm 1}$ this dispersion has a minimum at $k=0$  or $k=\pm \pi/2$. By continuity, we see that for dispersive phonons we obtain heavy but not infinite masses for nonzero $t$ at strong couplings. Contrast this result against that of the $t=0$ Breathing-Mode Peierls polaron with $\hat{O}^{\rm e}_i =  \tilde{\alpha} \hat{O}^{\rm BMP}_i = \tilde{\alpha} (\hat{c}_{i+1}^\dagger \hat{c}_{i} - \hat{c}_{i}^\dagger \hat{c}_{i-1} + {\rm h.c.})$ in Eq.~\eqref{Eq:Diagatomiclimit}. The different form of the electron operator means that one finds  a second nearest-neighbor hopping of the carrier: $+ \frac{\alpha^2}{\Omega} \sum_i (\hat{c}_{i+2}^\dagger \hat{c}_{i} + {\rm h.c.})$ even in the absence of phonon dispersion. This term produces a dispersion even in the atomic limit.  Importantly, in this limit the energy minimum is at $k=\pi/2$ (zone boundary of the Brillouin zone for second-neighbor hopping). At $\lambda = 0$, the energy minimum is at $k=0$, implying that the weak- and strong-coupling regimes must be separated by non-analyticity (phase transition), as seen in Fig.~\ref{fig:fig2}. We also verify this picture numerically in the atomic limit in Fig.~\ref{fig:fig3}, where we compute momentum-resolved spectral functions  for the models with dispersionless phonons. In the Bond-Peierls model, a phonon dispersion generates a second-neighbor hopping but the sign of the generated hopping amplitude may be either positive or negative, corresponding to a polaron band minimum at $k=\pm \pi/2$ or $k=0$ respectively, so that the weak- and strong-coupling regimes may or may not be separated by a transition, depending on the microscopic details.

 \begin{figure}[b]
    \includegraphics[width=\columnwidth]{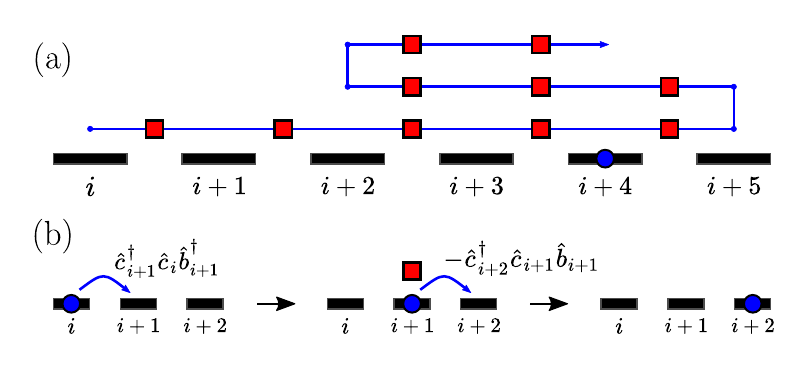}
    \caption{Cartoon depicting examples of virtual processes in which the carrier (blue circle) creates phonon excitations (red squares), forming a polaron. (a) A carrier in the Bond-Peierls (BP) model with dispersionless phonons moves from site $i\rightarrow i+1 \rightarrow i+2 \rightarrow ...$ and circles
    back multiple times (blue arrow) to form a state with 10 total
    phonons, terminating at site $i+4.$ We observe that there is no way for
    the carrier to move from site $i$ to some
    site $j \neq i$ whilst also annihilating all phonons on the lattice via $\hat{V}^{\rm BP}_{\rm e-ph}$,
    explaining the flat bands in Fig~\ref{fig:fig3}(a) in the $t=0$ limit. (b) A
    second-order phonon-mediated process  in
    the Breathing-Mode Peierls (BMP) model already permits second nearest-neighbor hopping (note the negative sign, responsible for the sharp transition \cite{marchand2010sharp}). We see that unlike in the Bond-Peierls model, the Breathing-Mode Peierls
    coupling $\hat{V}^{\rm BMP}_{\rm e-ph}$ directly mediates  carrier mobility.}
    \label{fig:figCartoon}
\end{figure}

The inverse mass of the Bond-Peierls polaron in the model with dispersionless phonons found in Fig.~\ref{fig:fig2}(f) empirically exhibits a dependence on $\lambda$ in the regimes we study that deviates from an exponential decrease. This apparent non-exponential dependence may be rationalized as follows.  The electronic portion of the canonically transformed Hamiltonian in the $t=0$ limit is just $-\frac{\tilde{\alpha}^2}{{\mathcal D}_0}\sum_i \hat{n}_i$, so in the transformed basis one obtains a degenerate family of ground states corresponding to transformed electrons localized  on sites $i$ with binding energy $\tilde{\alpha}^2/{\mathcal D}_0$. Undoing the transformation we see that in the original basis the ground state wavefunction is  $e^{i\frac{\tilde{\alpha}}{\mathcal{D}_0}\sum_j\hat{P}_j\hat{\mathcal{O}}^{\rm BP}_j}\ket{i}$ with $\ket{i}$ the wavefunction of a carrier localized on site $i$. The transformation factor delocalizes the electron over a distance that grows with $\alpha$. This physics may be understood as an analogue of the string picture familiar from analyses of the Ising limit of the $t-J$ model~\cite{trugman1988interaction,berciu2010momentum,Sous2020Fracton1,Sous2020Fracton2}, as follows. As the electron moves away from its original site, each  hop to a new site requires the creation of a phonon on the connecting bond. This can only be accomplished via a virtual process in which the electron ultimately retraces its path to the original site in order to absorb the bosons, see Fig.~\ref{fig:figCartoon}. The length of this string depends on the ratio $\alpha/\Omega$. Thus, as $\lambda$ increases the Bond-Peierls polaron becomes increasingly delocalized over many sites in sharp contrast to e.g. the Holstein model in which at $t=0$ the electron is strictly localized to one site.   The bare hopping $t$ connects the states (degenerate at $t=0$) in which the electron wavefunction is centered around sites $i$ and $i\pm 1$. The large spatial extent of the wavefunction in the strong-coupling limit means that the two states share very similar phonon configurations, explaining why the overlap between the two states is not exponentially small, unlike in the Holstein model. 

It is important to note that the arguments of the canonical transformation presented above carry over to higher dimensions and we thus  expect the qualitative features of  polarons and bipolarons in Bond- and Breathing-Mode Peierls systems, including the existence of a sharp polaronic transition and the behavior of the polaron and bipolaron masses as a function of $\lambda$  to qualitatively extend to higher dimensions.

\emph{Conclusions and outlook}.--- In this work, we studied the Bond-Peierls polaron in 1D, and contrasted its properties with the Breathing-Mode Peierls polaron. The mass of the Bond-Peierls polaron in 1D increases with increasing coupling strength and for dispersionless phonons diverges in the infinite coupling limit. The divergence may be understood as arsing from a phonon string-like mechanism, which is destroyed for any non-zero phonon dispersion. In the strong coupling limit of the bond-Peierls problem the dispersion minimum may be at $k=0$ or $k\neq 0$ depending on the momentum dependence of the phonons. A minimum at $k\neq0$ implies a phase transition at an intermediate value of the coupling. Our investigation poses interesting questions about whether the transition at nonzero temperatures, which may have relevance for real materials, especially in organic materials~\cite{McKenzieOrganics} and certain classes of perovskites~\cite{mayers2018lattice}. Our results demonstrate that Peierls polarons  exhibit non-exponential mass enhancement with $\lambda$ at least in the regimes of parameter space we can accurately simulate as well as possible transitions in presence of phonon dispersion characteristic of real systems, although the specific details depend on the specific model. This sensitivity to details 
suggest that some of the assumed phenomenology of Peierls materials~\cite{BoSSH2D,CaiAFM,Assaad2D} may require a  careful analysis of  universal versus model specific behavior.

Based on a canonical transformation derived in the atomic limit, we have argued that the qualitative features of polarons in Peierls models, including their sharp transition and mass dependence on the coupling strength, extend to higher dimensions.  Importantly, the relatively light effective masses of Bond-Peierls polarons when compared to Holstein polarons even for strong couplings opens the possibility that the model may generate light bipolarons in two dimensions. The mass of bipolarons generally depends on two important factors: 1) the single polaron mass, and 2) the nature of pairing mechanism. A kinetic-energy-enhancing pairing mechanism can in principle bind relatively heavy polarons~\cite{SousBipolaron,Sous2020Fracton2}, forming bipolarons that are not necessarily heavier. However, the Breathing-Mode Peierls model shows the most promise for producing a novel route to high temperature bipolaronic superconductivity as it should describe light bipolarons in two dimensions due to it ability to generate 1) light single polarons at strong couplings~\cite{marchand2010sharp}, 2) pair-hopping interactions, which mediate the formation of light bipolarons~\cite{SousBipolaron}.

\emph{Note added}.--- During the writing of this work, we became aware of a related study of Ref.~\onlinecite{Prokofiev2DPeierls}, which focuses on the (Breathing-Mode and Bond-)Peierls polarons in two dimensions. The main qualitative features of our results agree where they intersect. Specifically, the two works affirm the absence of exponential mass enhancement of the Bond-Peierls polaron in the regimes simulated. Furthermore, their agreement confirms that the behavior of Peierls polarons exhibit little dimensional dependence. 

\emph{Acknowledgements}.---  M.~R.~C. acknowledges the following support: This material is based upon work supported by the U.S. Department of Energy, Office of Science, Office of Advanced Scientific Computing Research, Department of Energy Computational Science Graduate Fellowship under Award Number DE-FG02-97ER25308. A.~J.~M., D.~R.~R. and J.~S. acknowledge support from the National Science Foundation (NSF) Materials Research Science and Engineering Centers (MRSEC) program through Columbia University in the Center for Precision Assembly of Superstratic and Superatomic Solids under Grant No. DMR-1420634. J.~S. also acknowledges the hospitality of the Center for Computational Quantum Physics (CCQ) at the Flatiron Institute. The Flatiron Institute is a division of the Simons Foundation.

\emph{Disclaimer}.--- This report was prepared as an account of work sponsored by an agency of the United States Government. Neither the United States Government nor any agency thereof, nor any of their employees, makes any warranty, express or implied, or assumes any legal liability or responsibility for the accuracy, completeness, or usefulness of any information, apparatus, product, or process disclosed, or represents that its use would not infringe privately owned rights. Reference herein to any specific commercial product, process, or service by trade name, trademark, manufacturer, or otherwise does not necessarily constitute or imply its endorsement, recommendation, or favoring by the United States Government or any agency thereof. The views and opinions of authors expressed herein do not necessarily state or reflect those of the United States Government or any agency thereof.

\providecommand{\noopsort}[1]{}\providecommand{\singleletter}[1]{#1}%

\end{document}